
\input amstex
\documentstyle{amsppt}
\magnification=1200
\NoBlackBoxes
\NoRunningHeads
\define\conf{\operatorname{conf}}
\define\ou{\text{\'o}}
\define\Dscr{\operatorname{Dscr}}
\define\slt{\operatorname{sl}(2,\Bbb C)}
\define\Eidos{\operatorname{E\widetilde{\iota}\delta o\varsigma}}
\define\Logos{\operatorname{\Lambda\ou\gamma o\varsigma}}
\define\Tropos{\operatorname{T\rho\ou\pi o\varsigma}}
\define\SU{\operatorname{SU}}
\define\Vrt{\operatorname{Vert}}
\define\Const{\operatorname{Const}}
\define\sgn{\operatorname{sgn}}
\define\V{W}
\define\jlt{\operatorname{jl}(2,\Bbb C)}
\define\slth{\operatorname{sl}(3,\Bbb C)}
\define\tr{\operatorname{tr}}
\redefine\t{\operatorname{T}}
\redefine\i{\operatorname{int}}
\define\e{\operatorname{ext}}
\define\pPhi{\partial_t\Phi}
\define\pu{\partial_t u}
\document
\qquad\qquad\qquad\qquad\qquad\qquad\qquad\qquad\qquad\qquad\qquad\qquad\qquad
hep-th/9401067
\centerline{}
\centerline{}
\centerline{}
\topmatter
\title
\centerline{VISUALIZING 2D QUANTUM FIELD THEORY:}
\centerline{GEOMETRY AND INFORMATICS OF MOBILEVISION}
\endtitle
\author
\centerline{\qquad\qquad\qquad\qquad\qquad\qquad
{\rm To the Bicentenary of N.I.Lobachevskii.}}
\centerline{}
\centerline{}
\centerline{{\tt D.JURIEV}}
\centerline{}
{\rm Mathematical Division,\linebreak Research Institute for System
Studies\linebreak
of Russian Academy of Sciences, ul.Avtozavodskaya 23,\linebreak
109280 Moscow, Rep. of Russia}\linebreak
\centerline{}
\endauthor
\date July, 1993\enddate
\abstract
This article is devoted to some interesting geometric and informatic
interpretations of peculiarities of 2D quantum field theory, which become
revealed after its visualization.
\centerline{}
\centerline{}
Electronic mail: juriev\@systud.msk.su\newline
\centerline{}
\endabstract
\endtopmatter
\eightpoint
The present research note, being addressed to specialists in the modern
quantum field theory as well as in the computer graphics and the scientific
visualization, is an attempt (maybe sometimes only suggestive or rather
discussional) to describe some peculiarities of two--dimensional quantum field
theory becoming revealed after its visualization. The problem of visualization
is considered from both geometric and informatic points of view, which are in
some sense complementary to each other. Our crucial example is Mobilevision --
a certain "anomalous virtual reality" "naturalizing" 2D
quantum projective ($\slt$--invariant) field theory (precise definitions of
the concepts of anomalous virtual reality and the naturalization see below);
so the theme of the article is geometry and informatics of Mobilevision.

The text is organized as follows; it includes two paragraphs: in the first one
the visualized peculiarities of 2D quantum field theory are described in
geometric terms, in the second -- in informatic ones; of course, both
paragraphs are interrelated by common constructions (although considered from
rather different points of view) in accord with the thought that "Die
Dualit\"at der $\Eidos$ und $\Logos$ wird aufgehoben in ihr $\Tropos$".

\head 1. GEOMETRY OF MOBILEVISION\endhead

It is well-known that many geometric discoveries were motivated by preceeding
exploration of various natural phenomena. It seems that not less interesting
geometric constructions may appear after investigations of artificial
(man--made) systems; especially, of computer ones, which are partially based
on modelling of a class of the most intriguing natural phenomena.

\subhead 1.1. Interpretational geometry and anomalous virtual realities {\rm
[1]} \endsubhead

Interpretational geometry is a certain geometry related to
interactive computerographical psychoinformation systems.
Mathematical data in such systems exist in the form of an
interrelation of an interior geometric image (figure) in the
subjective space of observer and an exterior computerographical
representation; the least includes the visible
elements (draws of figure) as well as of the invisible ones
(f.e. analytic expressions and algorythms of the constructing of such draws).
Process of the corresponding of a geometrical image (figure) in the interior
space of observer to a computerographical representation (visible and
invisible elements) is called {\it translation};
the visible object maybe nonidentical to the
figure, f.e. if a 3--dimensional body is defined by an axonometry, in three
projections, cross-sections or cuts, or in the window technique, which allows
to scale up a concrete detail of a draw (which is a rather pithy operation for
the visualization of fractals), etc, in this case partial visible
elemnts may be regarded as modules, which translation is realized separately;
the translation is called by {\it interpretation\/} if the translation of
partial modules is
realized depending on the result of the translation of preceeding ones.

\definition{Definition 1} A figure, which is obtained as a result of the
interpretation, is called {\it interpretational figure\/}; the draw of an
interpretational figure is called {\it symbolic\/}.\enddefinition

The computer--geometric description of mathematical data in interactive
information systems is deeply related to the concept of anomalous virtual
reality.
It should be mentioned that there exist several approaches to
foundations of geometry:
in one of them the basic geometric concept is a space (a medium, a field),
geometry describes various properties of a space and its states, which are
called the draws of figures;
it is convenient to follow this approach for the purposes of the describing of
geometry of interactive information systems; the role of the medium
is played by an anomalous virtual reality, the draws of figures are its
certain states.

\definition{Definition 2}

A. {\it Anomalous virtual reality\/} ({\it AVR\/}) {\it in a narrow sense\/}
is a certain system of rules of non--standard descriptive geometry adopted to
a realization on videocomputer (or multisensor system of "virtual reality"
[2,3]);
{\it anomalous virtual reality in a wide sense\/} contains also an image in
the cyberspace made accordingly to such system of rules;
we shall use this term in a narrow sense below.

B. {\it Naturalization\/} is the corresponding of an AVR
to an abstract geometry or a physical model;
we shall say that the AVR {\it naturalizes\/} the model
and such model {\it transcendizes\/} the naturalizing AVR.
{\it Visualization in a narrow sense\/} is the corresponding of certain images
or visual dynamics in the AVR to objects of the abstract geometry
or processes in the physical model; {\it visualization in a wide sense\/} also
includes the preceeding naturalization.

C. An anomalous virtual reality, whose images depends on an observer, is
called {\it intentional anomalous virtual reality\/} ({\it IAVR\/});
generalized perspective laws in IAVR contain the equations of dynamics of
observed images besides standard (geometric) perspective laws; a process of
observation in IAVR contains a physical process of observation and a virtual
process of intention, which directs an evolution of images accordingly to
dynamical laws of perspective.
\enddefinition

In the intentional anomalous virtual reality objects of observation present
themselves being connected with observer, who acting on them in some way,
determines, fixes their observed states, so an object is thought as a
potentiality of a state from the defined spectrum, but its realization depends
also on observer;
the symbolic draws of interpretational figures are presented by states of a
certain IAVR.

It should be mentioned that a difference of descriptive geometry of
computerographical information systems from the classical one is the presense
of colors as important bearers of visual information;
a reduction to shape graphics, which is adopted in standard descriptive
geometry, is very inconvenient, since the use of colors is very familiar in
the scientific visualization [4--6].
The approach to the computerographical interactive information systems based
on the concept of anomalous virtual reality allows to consider an
investigation of structure of a color space as a rather pithy problem of
descriptive geometry, because such space maybe much larger than the usual one
and its structure may be rather complicated.
\pagebreak

\definition{Definition 2D}
A set of continuously distributed visual characteristics of image in an
anomalous virtual reality is called {\it anomalous color space\/}; elements
of an anomalous color space, which have non--color nature, are called {\it
overcolors\/}, and quantities, which transcendize them in the abstract model,
are called {\it "latent lights"\/}. {\it Color--perspective system\/} is a
fixed set of generalized perspective laws in fixed anomalous color space; two
AVRs are called equivalent iff their color--perspective systems coincide; an
AVR, which is equivalent to one realized on the videocomputer, is called {\it
marginal}.
\enddefinition

\subhead 1.2. Quantum projective field theory and Mobilevision {\rm [1,7,8]}
\endsubhead

It seems to be a significant fact that 2D quantum field theory maybe expressed
in terms of interpretational geometry, so that various objects of this theory
are represented by interpretational figures. Our crucial example is
{\it Mobilevision\/} ({\it MV\/}).
MV is an IAVR naturalizing {\it the
quantum projective field theory\/} ({\it QPFT\/}; [8] and refs wherein);
the process of naturalization is described in [7,1]; its key points
will be presented below, here our attention is concentrated on
the basic concepts of the
QPFT, which naturalization Mobilevision is.

\definition{Definition 3A} {\it QFT--operator algebra\/} ({\it
operator algebra of the quantum field theory, vertex operator algebra, vertex
algebra\/}) is the pair $(H,t^k_{ij}(\vec x))$: $H$ is a linear space,
$t^k_{ij}(\vec x)$ is $H$--valued tensor field such that $t^l_{im}(\vec
x)t^m_{jk}(\vec y)=t^m_{ij}(\vec x-\vec y)t^l_{mk}(\vec y)$.
\enddefinition

Let us intruduce the operators $l_{\vec x}(e_i)e_j=t^k_{ij}(\vec x)e_k$, then
the following relations will hold: $l_{\vec x}(e_i)l_{\vec
y}(e_j)=t^k_{ij}(\vec x-\vec y)l_{\vec y}(e_k)$ ({\it operator product
expansion\/}) and $l_{\vec x}(e_i)l_{\vec y}(e_j)=l_{\vec y}(l_{\vec x-\vec
y}(e_i)e_j)$ ({\it duality\/}).
Also an arbitrary QFT--operator algebra one can define an operation depending
on the parameter: $m_{\vec x}(e_i,e_j)=t^k_{ij}(\vec x)e_k$; for this
operation the following identity holds: $m_{\vec x}(\cdot,m_{\vec
y}(\cdot,\cdot))=m_{\vec y}(m_{\vec x-\vec y}(\cdot,\cdot),\cdot)$; the
operators $l_{\vec x}(f)$ are the operators of the left multiplication in the
obtained algebra.

\definition{Definition 3B}
QFT--operator algebra $(H,t^k_{ij}(u); u\in\Bbb C)$ is called {\it (derived)
QPFT--operator
algebra\/}
iff (1) $H$ is the sum of Verma modules $V_{\alpha}$ over $\slt$ with the
highest vectors $v_{\alpha}$ and the highest weights $h_{\alpha}$, (2)
$l_u(v_{\alpha})$ is a primary field of spin $h_{\alpha}$, i.e.
$[L_k,l_u(v_\alpha)]=(-u)^k(u\partial_u+(k+1)h_{\alpha})l_u(v_{\alpha})$,
where $L_k$ are the $\slt$ generators ($[L_i,L_j]=(i-j)L_{i+j}$,
$i,j=-1,0,1$),
(3) the (derived) rule of descendants generation holds
($[L_{-1}l_u(f)]=l_u(L_{-1}f)$).
(Derived) QPFT--operator algebra $(H,t^k_{ij}(u))$ is
called {\it projective $G$--hypermultiplet\/}, iff the group $G$ acts in it by
automorphisms, otherwords, the space $H$ possesses a structure of the
representation of the group $G$, the representation operators commute with the
action of $\slt$ and $l_u(T(g)f)=T(g)l_u(f)T(g^{-1})$.
\enddefinition

The linear spaces of the highest vectors of the fixed weight form
subrepresentations of $G$, which are called {\it multiplets\/} of projective
$G$-hypermultiplet.

As it was mentioned above MV is a certain
AVR, which naturalizes the QPFT.
Possibly, MV is not its unique naturalization; here we describe
the key moments of the process of naturalization of the
QPFT which is resulted in MV.
Unless the abstract model (QPFT) has a quantum
character the images in its naturalization
(MV) are classical;
the transition from the quantum field model to classical one is done by
standard rules [16], namely, the classical field with Taylor coefficients
$|a_k|^2$ is corresponded to the element $\sum a_k L_{-1}^k v_{\alpha}$ of the
QPFT--operator algebra.
Under the naturalization three classical fields are identified with fields of
three basic colors (red, green and blue), other fields with
fields of overcolors; there are pictured only the color characteristics for
the fixed moment of time on the screen of the videocomputer as well as the
perception of the overcolors by an observer is determined by the intentional
character of the AVR of Mobilevision.
Namely, during the process of the evolution of the image, produced by the
observation, the vacillations of the color fields take place in accordance to
the dynamical perspective laws of
MV (Euler formulas or Euler--Arnold equations).
These vacillations depend on the character of an observation (f.e. an eye
movement or another dynamical parameters); the vacillating image depends on
the distribution of the overcolors, that allows to interpret the overcolors
as certain vacillations of the ordinary colors.
So the overcolors of
MV are vacillations of the fixed type and structure of ordinary
colors with the defined dependence on the parameters of the observation
process;
the transcending "latent lights" are the quantized fields of the basic model
of the QPFT.
\pagebreak

The presence of the $\SU(3)$--symmetry of classical color
space
allows to suppose that the QPFT--operator algebra of the initial model is the
projective $\SU(3)$--hypermultiplet.

\subhead 1.3. Quantum conformal and $q_R$--conformal field theories;
quantum--field analogs of Euler--Arnold tops {\rm [1,8]}
\endsubhead

\definition{Definition 4A {\rm [9-11]}}
The highest vector $T$ of the weight 2 in the QPFT--opeartor algebra will
be called {\it the conformal stress--energy tensor\/} if
$T(u):=l_u(T)=\sum L_k
(-u)^{k-2}$, where the operators $L_k$ form the Virasoro algebra:
$[L_i,L_j]=(i-j)L_{i+j}+\frac{i^3-i}{12} c\cdot I$.
The set of the highest vectors $J^\alpha$ of the weight 1 in the
QPFT--operator algebra will be called the set of {\it the affine currents\/}
if $J^\alpha(u):=l_u(J^\alpha)=\sum J^\alpha_k(-u)^{k-1}$, where the
operators $J^\alpha_k$ form {\it the affine Lie algebra\/}:
$[J^\alpha_i,J^\beta_j]=c^{\alpha\beta}_\gamma
J^\gamma_{i+j}+k^{\alpha\beta}\cdot i\delta(i+j)\cdot I$.
\enddefinition

If there is defined a set of the affine currents in the QPFT--operator algebra
then one can construct the conformal stress--energy tensor by use of
Sugawara construction or more generally by use of the Virasoro
master equations.
Below we shall be interested in the special deformations of the quantum
conformal field theories (namely, ones with central charge $c=4$)
in class of the quantum projective ones, which will be called
{\it quantum $q_R$--conformal field theories\/}; the crucial role is played by
so--called {\it Lobachevskii algebra\/} in their constructions.
In the Poincare realization of the Lobachevskii plane (the realization in the
unit disk) the Lobachevskii metric maybe written as
$w=q_R^{-1}\,dzd\bar{z}/(1-|z|^2)^2$;
one can construct the $C^*$--algebra (Lobachevskii algebra), which maybe
considered as a quantization of such metric, namely, let us consider
two variables $t$ and $t^*$, which obey the following commutation relations:
$[tt^*,t^*t]=0$, $[t,t^*]=q_R(1-tt^*)(1-t^*t)$ (or in an equivalent form
$[ss^*,s^*s]=0$, $[s,s^*]=(1-q_Rss^*)(1-q_Rs^*s)$, where $s=(q_R)^{-1/2}t$);
one may realize such variables by bounded
operators in the Verma module over $\slt$ of the weight
$h=\frac{q_R^{-1}+1}2$;
if such Verma module is realized in polynomials of one complex variable $z$
and the action of $\slt$ has the form $L_{-1}=z$, $L_0=z\partial_z+h$,
$L_1=z(\partial_z)^2+2h\partial_z$, then the variables $t$ and $t^*$ are
represented by tensor operators $D=\partial_z$ and $F=z/(z\partial_z+2h)$.
These operators are bounded if
$q_R>0$ and therefore one can construct a Banach algebra
generated by them and obeying the prescribed commutation relations;
the structure of $C^*$--algebra is rather obvious: an involution $*$ is
defined on generators in a natural way, because the corresponding tensor
operators are conjugate to each other.

\definition{Definition 4B}
The highest vector $T$ of the weight 2 in the QPFT--operator algebra will
be called {\it the $q_R$--conformal stress--energy tensor\/} if
$T(u):=l_u(T)=\sum L_k (-u)^{k-2}$, where the operators $L_k$ form the
$q_R$--Virasoro algebra: $[L_i,L_j]=(i-j)L_{i+j}$ ($i,j\ge -1$; $i,j\le 1$),
$[L_2,L_{-2}]=H(L_0+1)-H(L_0-1)$,
$H(t)=t(t+1)(t+3h-1)^2/((t+2h)(t+2h-1))$,
$h=(q_R^{-1}+1)/2$ (cf.[12]).
The set of the highest vectors $J^\alpha$ of the weight 1 in the
QPFT--operator algebra will be called the set of {\it the $q_R$--affine
currents\/} if $J^\alpha(u):=l_u(J^\alpha)=\sum J^\alpha_k(-u)^{k-1}$, where
the operators $J^\alpha_k$ form {\it the $q_R$--affine Lie algebra\/}:
$J^{\alpha}_k=J^{\alpha}T^{-k}f_k(t)$,
$[J^{\alpha},J^{\beta}]=c^{\alpha\beta}_{\gamma}J^{\gamma}$,
$Tf(t)=f(t+1)T$,
$[T,J^{\alpha}]=[f(t),J^{\alpha}]=0$,
$f_k(t)=t\ldots(t-k),\text{ if } k\ge 0,\text{ and }
((t+2h)\ldots (t+2h-k+1))^{-1},\text{ if } k\le 0$,
$h=(q^{-1}_R+1)/2$.
\enddefinition

It should be mentioned that $q_R$--affine currents and $q_R$--conformal
stress--energy tensor are just the $\slt$--primary fields in the Verma module
$V_h$ ($h=\frac{q^{-1}_R+1}2$) over $\slt$ of spin 1 and 2, respectively;
if such module is realized as before then
$J_k=\partial_z^k$, $J_{-k}=z^k/(\xi+2h)\ldots(\xi+2h+k-1)$;
$L_2=(\xi+3h)\partial^2_z$, $L_1=(\xi+2h)\partial_z$, $L_0=\xi+h$,
$L_{-1}=z$,
$L_{-2}=z^2\frac{\xi+3h}/{(\xi+2h)(\xi+2h+1)}$, $\xi=z\partial_z$.
So the generators $J_k^\alpha$ of $q_R$--affine algebra maybe represented via
generators of Lobachevskii
$C^*$--algebra:
$J^\alpha_k=J^\alpha t^k,\text{ if } k\ge 0,\text{ and }
J^\alpha(t^*)^{-k},\text{ if } k\le 0$,
($[J^\alpha,J^\beta]=c^{\alpha,\beta}_{\gamma}J^{\gamma}$).
That means that $q_R$--affine algebra admits a homomorphism in a tensor
product of the universal envelopping algebra $\Cal U(\frak g)$ of the Lie
algebra $\frak g$, generated by $J^\alpha$, and Lobachevskii algebra.
The (derived) QPFT--operator algebras generated by $q_R$--affine currents are
called canonical projective $G$--hypermultiplets.
The primary fields $V_k(u)=\exp(k(Q+R(\int V_1(u)\,du)))$
($R(u^n)=-\sgn(n)u^n$, i.e. $R$ is the Hilbert transform
$f(\exp(it))\mapsto-\frac{i}{2\pi}\int f(\exp(i(t-s)))\cot(s/2)\,ds$; a charge
$Q$ is defined as $Q(z^n)=\sum_{j=0}^{n-1}(j+2h)^{-1}z^n$; [13,14])
\footnote{ It should be mentioned that the operator field $V_\mu(u)$ maybe
factorized as $\Const\cdot\exp(\mu\psi_+(u))\cdot\mathbreak A_\mu
u^{-\mu}\exp(\mu\psi_-(u))$,
where $\psi_+(u)$ and $\psi_-(u)$ are regular and singular parts of $R(\int
V_1(u)\,du)$, respectively: $\psi_+(u)=-\log(1+uF)$,
$\psi_-(u)=-\log(1+u^{-1}D)$; the operator $A_\mu$ is defined as $A_\mu=\sum
\Gamma(k-\mu)F^kD^k/k!=
\Const\cdot\Gamma(z\partial_z+2h-\mu)/\Gamma(z\partial_z+2h)$,
$Q=\frac{d}{d\mu}\left.A_\mu\right|_{\mu=0}+\Const$.} of non--negative integer
spins $k$ in the Verma module $V_h$, which form a closed QPFT--operator
algebra (a subalgebra of $\Vrt(\slt)$ [17,18], generated by vertex operator
fields $B_k(u;\nabla_h)$ [16,17]), are not mutually local.
Nevertheless, their commutation relations may be described as follows
$V_\alpha(u)V_\beta(v)=
S^{\gamma\delta}_{\alpha\beta}(u-v)V_\gamma(v)V_\delta(u)$,
that means that these primary fields form a Zamolodchikov algebra with
$S$--matrix
$S^{\gamma\delta}_{\alpha\beta}(u;q_R)=
u^{-\alpha-\beta+\gamma+\delta}S^{\gamma\delta}_{\alpha\beta}(q_R)$;
$S^{\gamma\delta}_{\alpha\beta}(q_R)=0$ if $\alpha+\gamma\ne\beta+\delta$.
It is interesting to calculate $T$--exponent and monodromy of $q_R$--affine
current; it maybe easily performed by a perturbation of simple formulas for
such objects for a singular part of a current, as it was stated in [7] such
perturbation by a regular part does not change the resulting monodromy.

Let $H$ be an arbitrary direct sum of Verma modules over $\slt$ and $P$ be a
trivial fiber bundle over $\Bbb C$ with fibers isomorphic to $H$;
it should be mentioned that $P$ is naturally trivialized and possesses a
structure of $\slt$--homogeneous bundle.
A $\slt$--invariant Finsler connection $A(u,\partial_t u)$ in $P$ is called an
angular field;
angular field $A(u,\partial_t u)$ may be expanded by $(\partial_t u)^k$, the
coefficients of such expansion are just $\slt$--primary fields;
the equation
$\partial_t\Phi_t=A(u,\partial_t u)\Phi_t$,
where $\Phi_t$ belongs to $H$ and $u=u(t)$ is the function of scanning, is a
quantum--field analog of the Euler formulas;
such analog describes an evolution of MV image under the
observation (more rigorously, such evolution is defined in the dual space
$H^*$ by formulas $\partial_t\Phi_t=A^{\t}(u,\partial_t u)\Phi_t$).
Regarding canonical projective $G$--hypermultiplet we may construct a quantum
field analog of the Euler--Arnold equation
$\partial_t A=\{H,A_t\}$,
where an angular field $A(u,\partial_t u)$ is considered as an element of the
canonical projective $G$--hypermultiplet being expanded by $\slt$--primary
fields of this hypermultiplet, $H$ is the quadratic element of $S(\frak g)$,
$\{\cdot,\cdot\}$ are canonical Poisson brackets in $S(\frak g)$.
It is quite natural to demand $H$ be a solution of the Virasoro master
equation.
If we consider a projective $G$--hypermultiplet, which is a semi--direct
product of the canonical one and a trivial one (i.e. with $l_u(f)=0$), it
will be possible to combine Euler--Arnold equations with Euler formulas to
receive the complete dynamical perspective laws of the MV.

\subhead 1.4. Organizing MV cyberspace \endsubhead

MV cyberspace consists of a space of images $V_I$ with the fundamental length
(a step of the lattice) $\Delta_I$ and a space of observation $V_O$ with the
fundamental length $\Delta_O$; the space of images $V_I$ is one where pictures
are formed, whereas the space of observation $V_O$ is used for a detection of
motion of eyes; it is natural to claim that $\Delta_O<\!<A_{\tr}$, where
$A_{\tr}$ is an amplitude of the eye tremor, as well as
$\Delta_I>\!>\Delta_O$.
The Euler formulas maybe written as $\partial_t\Phi_t=A(u,\partial_t
u)\Phi_t$, $A(u,\partial_t u)$ maybe considered approximately in the form
$M_1(t)\partial_t uV_1(u)+M_2(t)(\partial_t u)^2 V_2(u)+M_3(t)(\partial_t u)^3
V_3(u)$, where $\Phi_t\in H$ (or in the form
$\partial_t\Phi_t=A^{\t}(u,\partial_t u)\Phi_t$, $\Phi_t\in H^*$); here
$M_i(t)$
are data of Euler--Arnold top, $V_i(u)$ are $\slt$--primary fields in the Verma
module $V_h$, which maybe written as
$V_i(u)=(-u)^{-i}(\V_i(u)+\V^*_i(u)-D^i_0)=\sum_{j\in\Bbb Z}
(-u)^{-i-j}D^i_j$, $\V_i(u)=\sum_{j\ge 0} (-u)^{-j}D^i_j$,
$\V^*_i(u)=\sum_{j\ge 0} (-u)^jD^i_{-j}$, $D^i_{-j}=(D^i_j)^*$, where $*$ is
the conjugation in the unitarizable Verma module.
The tensor operators $D^i_k$ ($k\ge 0$, $i=1,2,3$) have the form
$D^i_k=P_{i,k}(z\partial_z)\partial^k_z$, $P_{1,k}(t)=1$,
$P_{2,k}(t)=t+(k+1)h$, $P_{3,k}(t)=t^2+((k+2)h+k/2)t+h(2h+1)(k+1)(k+2)/6$.
It should be mentioned that the fields $\V^{\t}_i(u)$ in local $\slt$--modules
$V^*_h$ are defined by rather simple expressions:
$$
\align
\V^{\t}_1(u)=&\frac1{1-u^{-1}x},\\
\V^{\t}_2(u)=&-\frac{x}{1-u^{-1}x}\partial_x+\frac{h}{(1-u^{-1}x)^2},\\
\V^{\t}_3(u)=
&\frac{x^2}{1-u^{-1}x}\partial^2_x-(2h+1)\frac{x}{(1-u^{-1}x)^2}\partial_x+
\frac{h(2h+1)}{3}\frac{1}{(1-u^{-1}x)^3}.
\endalign
$$
The matrix $A^{\t}(u,\partial_t u)$ of size $(N,N)$ ($N$ is a number of points
of $V_I$) should be expanded in a sum of three terms $M_i(t)(\partial_t
u)^iV_i(u)$ ($i=1,2,3$), where $V^{\t}_i(u)$ are matrices of size $(N,N)$,
depending on parameter $u$; this parameter may have $M$ different values ($M$
is number of points of $V_O$).
Matrices $\V^{\t}_i(u)$ are easily calculated, one should obtain the complete
matrices $V^{\t}_i(u)$ making a conjugation in the unitary local
$\slt$--module $V^*_h$.
Derivatives should be replaced by differences everywhere in a standard way.
Formulas for $M_i(t)$ maybe received from [19].

\subhead 1.5. Non--Alexandrian geometry of Mobilevision \endsubhead

It should be marked that almost all classical geometries use a certain
postulate, which we shall call Alexandrian, but do not include it in their
axiomatics explicitely.
For our purposes we prefer to give a precise formulation of this postulate.

\proclaim{Alexandrian postulate} Any statement holding for a certain geometric
configuration remains true if this configuration is considered as a
subconfiguration of any its extension.
\endproclaim

Alexandrian postulate means that an addition of any subsidiary objects to a
given geometric configuration does not influence on this configuration.
It is convenient to describe a well--known example of non--Alexandrian
geometry (which maybe called Einstein geometry).

\remark{Example of non-Alexandrian geometry} Objects of geometry are weighted
points and lines.
Weigh-\linebreak ted points are pairs (a standard point on a plane, a real
number). They
define a (singular) metric on a plane via Einstein--type equations $R(x)=\sum
m_\alpha\delta(x-x_\alpha)$, where $(x_\alpha,m_\alpha)$ are weighted points
and $R(x)$ is a scalar curvature.
Lines are geodesics for this metric.
The basic relation is a relation of an incidence.
\endremark

It can be easily shown taht Alexandrian postulate doesn't hold for such
geometry, which contains a standard Euclidean one (extracted by the condition
that all "masses" $m_\alpha$ are equal to 0).

Kinematics and process of scattering of figures maybe illustrated by another
important example of non--Alexandrian geometry --- geometry of solitons
[20,21].
The basic objects of KdV--soliton geometry are moving points on a line; a
configuration of such points defines a $n$--soliton solution of KdV--equation
$u_t=6uu_{xx}-u_{xxx}$ by the formulas $u(x,t)=-2(\log(\det(E+C)))_{xx}$,
where
$C_{nm}=
c_n(t)c_m(t)\exp(-(\varkappa_n+\varkappa_m)t)/(\varkappa_n+\varkappa_m)$,
$c_n(t)=c_n(0)\exp(4\varkappa^3_nt)$; such solution is asymptotically free,
i.e. maybe represented as a sum of 1--soliton solutions (solitons) whereas
$t\to\pm\infty$.
Soliton has the form
$u(x,t)=-2\varkappa^2\cosh^{-2}(\varkappa(x-4\varkappa^2t-\varphi))$, where
phase $\varphi$ is an initial position of soliton and $v=4\varkappa^2$ is its
velocity; scattering of solitons is two--particle, the shift of phases is
equal to
$\varkappa^{-1}_1\log|(\varkappa_1+\varkappa_2)|/|(\varkappa_1-\varkappa_2)|$
for the first (quick) soliton and
$-\varkappa_2^{-1}\log|(\varkappa_1+\varkappa_2)|/|(\varkappa_1-\varkappa_2)|$
for the second (slow) one.
Analogous scheme maybe realised for Sine--Gordon equation, which has the form
$u_{\xi\xi}-u_{\tau\tau}+\sin u=0$ in the light--cone variables; SG--soliton
is defined by the expression
$u(\xi,\tau)=4\arctan(\exp(-(1-v^2)^{1/2}(\xi-\xi_0-v\tau)))$; the shift of
the phases under the scattering is equal to $2(1-v^2)^{1/2}\log(v)$ for the
slow soliton (and by such expression with the opposite sign for the quick one)
in the coordinates of the mass centre, where $v$ is a relative velocity of the
quick soliton.
Another example of soliton geometry is connected with nonlinear Schr\"odinger
equation [22].
All examples of soliton geometries confirm the opinion that a breaking of the
Alexandrian postulate is generated by an interaction of geometrical objects,
in particular, such interaction maybe defined by a nonlinear character of
their evolution.

Let's consider now an interpretational scattering.
As it was stated below a figure in interpretational geometry is described by a
pair $(\Phi^{\i},\Phi^{\e})$, where $\Phi^{\i}$ is an interior image in the
subjective space of observer and $\Phi^{\e}$ is its exterior
computerographical draw; $\Phi^{\i}$ is a result of interpretation of
$\Phi^{\e}$.
It is natural to suppose that $\Phi^{\i}$ depends on $\Phi^{\e}$ functionally
$\Phi^{\i}_t=\Phi^{\i}\left[\Phi^{\e}_{\tau\le t}\right]$ and as a rule
nonlinearly; moreover, if $\Phi^{\e}$ is asymptotically free then $\Phi^{\i}$
is also asymptotically free.
Thus, a nontrivial scattering of interacting interpretational figures exists
(i.e. although we do not know an explicit form of dynamical equations for
$\Phi^{\i}$, their solutions, nevertheless, in view of our assumptions maybe
considered as soliton--like (!)), so interpretational geometries maybe
considered as non--Alexandrian ones; it should be specially marked that the
breaking of Alexandrian postulate is realized on the level of figures
themselves, but it is not observed on one of their draws.
\newpage

\head 2. INFORMATICS OF MOBILEVISION \endhead

\subhead 2.1. Information transmission via anomalous virtual realities:
AVR--photodosy
\endsubhead

As it was mentioned above informatics maybe considered as a point of view on
mathematical objects complementary to geometric one; so it seems to be useful
to reformulate our main definitions in informatics terms.
The concepts of AVR--photodosy and its formal grammatics are a natural
parallel to ones of anomalous virtual reality and the transcendizing abstract
model.

\definition{Definition 5A} The transmission of information via anomalous
virtual reality by "latent lights" is called {\it AVR--photodosy\/}; the
system of algebraic structures of the initial abstract model, which
characterizes a process of AVR--photodosy via naturalization, is called {\it
the formal grammatics\/} of AVR--photodosy.
\enddefinition

It should be mentioned that the concept of AVR--photodosy and its formal
grammatics is deeply related to one of anomalous color space, since the using
of such spaces allows to transmit diverse information in different forms, and
an investigation of the information transmission via AVRs, which character
depends on a structure of color space, is an important mathematical problem
(cf.[23]); it should be marked now that the results of such researches maybe
used for elaboration of computer systems of psychophysiological
self--regulation, hypnosis and suggestion.
Structure of AVR--photodosy is determined by its formal grammatics; in view of
the quantum character of formal grammatics of MV, AVR--photodosy via it is
quantum--logical [24]; it seems that this fact needs a special attention.
A few words more about MV formal grammatics: the investigation of a certain
systems of algebraic structures of the quantum projective field theory, which
maybe considered as possible candidates on the role of formal grammatics of MV
now, was begun by the author and his collaborators in various directions in
series of papers devoted to QPFT and is intensively continued in present time;
such investigations are far from a finish and there are a lot of hopes that
many new interesting algebraic objects will be found in the nearest future; it
should be marked that the state--of--the--art is very intriguing yet, the main
directions of researches are (1) the model formalism [25,15-17,26], (2) the
formalism of projective Fubini--Veneziano fields [13,14], (3) the formalism of
local field algebras [25,26], (4) a search for hidden symmetries [27-30], (4)
the commutative exterior differential calculus [31,32]; this article is not a
suitable place to describe the obtained results, a reader should address to
the cited list of references, especially to [25].

\subhead 2.2. Information transmission via intentional anomalous virtual
realities: IAVR--teleaesthesy
\endsubhead

At first, it should be mentioned that any IAVR is polysemantic, it means that
quantity and structure of an information, received by AVR--photodosy via it,
depends on observer; so it is a natural problem to describe the informatics of
interactive computerographical psychoinformation systems, which contain more
than one observer, in particular, a correlation of different observations.
Such systems maybe regarded as realizing an interactive MISD
(Multiple\_Instruction--Single\_Data) architecture with parallel interpretation
processes for different observers (this fact should be considered in a context
of a remark on a quantum--logical character of AVR--photodosy below); in this
way we encounter with the phenomenon, which is specific for such systems but
may have a more general meaning: namely, the processes of observation by
different observers induce a certain information exchange between them.

\definition{Definition 5B} AVR--photodosy via IAVR from one observer to
another is called {\it IAVR--teleaes-\linebreak thesy\/}; if during
IAVR--teleaesthesy
AVR--photodosies from different observers do not satisfy the superposition
principle then it will be said that {\it the collective effect\/} in
IAVR--teleaesthesy holds.
\enddefinition

It should be mentioned that (1) the process of IAVR--teleaesthesy has a
bilateral character; the observers entering into a teleaesthetic
communication, are simultaneously as inductors (observers sending information)
as receptors (ones receipting it), moreover, as it was mentioned below,
quantity and structure of recipted information depends on receptor as well as
on inductor; (2) the collective effect in IAVR--teleaesthesy means that the
different inductors in IAVR does not perceived as independent --- the
transmitted information is not a sum of informations sent by separate
observers, so the partial streams of information enter into the exchange
interaction, which form a specific information receipted by receptor.
A relation of an origin of IAVR--teleaesthesy to the fact that interactive
computerographical psychoinformation systems realize an interactive MISD
architecture should be specially marked.

\definition{Definition 5C}
An observer in a marginal AVR, to which there is no corresponded any observer
in the AVR, realized on the videocomputer, is called {\it virtual\/}; a
virtual observer, whose process of observation depends on several real
observers is called {\it a collective virtual observer}.
\enddefinition

The presence of a virtual observer means that a fixed part of the receipted
information is interpreted as an information sent by this observer, which does
not exist really.
The presence of a collective virtual observer is not obligatory but is usual
for interactive computerographical systems in the multi--user mode; this fact
also should be considered in a context of the previous remark that such
systems realize an interactive MISD architecture with parallel interpretation
processes for different users.

Let us now illustrate our constructions on the example of MV.
It should be mentioned that $q_R$--conformal field theory is a certain
deformation of $c=4$ conformal one; this deformation is infinitesimally
defined by the next Poisson brackets on the virasoro algebra with $c=4$:
$\{L_i,L_j\}=0$ ($i,j\ge -1$; $i,j\le 1$), $\{L_2,L_{-2}\}=R(L_0)$, where
$R(t)=2(2t^2+2t+1)$.
The initial quantum conformal field theory is local (the locality means that
$[l_u(X),l_v(Y)]=0$ if $u\ne y$) so the angular field $A^{\conf}(u_1,\ldots
u_n;\pu_1,\ldots\pu_n)$ corresponding to several observers, being a sum of
mutually local partial angular fields $A^{\conf}(u_i,\pu_i)$, defines a
MV--image dynamics obeying the superposition principle; so in this case a
collective effect in IAVR--teleaesthesy is not observed and a collective
virtual observer does not exist.
Let us now consider $q_R$--conformal field theory with non--zero $q_R$: the
corresponding partial angular fields are not mutually local (and therefore, a
collective effect of IAVR--teleaesthesy holds in the corresponding MV
anomalous virtual reality); nevertheless, they maybe expanded as
$A(u_i,\pu_i;q_R)=A^{\conf}(u_i,\pu_i)+q_R\delta A(u_i,\pu_i;q_R)$, where
$[A^{\conf}(u_i,\pu_i),A^{\conf}(u_j,\pu_j)]=0$ (so $q_R$ is a measure of the
collective effect); the multi--observer Euler formulas being of the form
$\pPhi_t=A(u_1,\ldots u_n;\pu_1,\ldots\pu_n;q_R)\Phi_t$ (where $A(u_1,\ldots
u_n,\pu_1,\ldots\pu_n;q_R)=\sum A(u_i,\pu_i;q_R)$) may be written as
$\pPhi_t=(\sum A^{\conf}(u_i,\pu_i)+q_R\delta A(u_1,\ldots
u_n;\pu_1,\ldots\pu_n;q_R))\Phi_t$ (where $\delta A(u_1,\ldots
u_n;\pu_1,\ldots u_n;q_R)=\sum\delta A(u_i,\pu_i;q_R)$).
These equations maybe considered as dynamical equations in a marginal IAVR,
real observers in which are described by conformal (mutually local) angular
fields $A^{\conf}(u_i,\pu_i)$ and a collective virtual observer is described
by the angular field $q_R\delta A(u_1,\ldots u_n;\pu_1,\ldots\pu_n;q_R)$; the
process of observation by a collective virtual observer is reduced to a
virtual process of intention, the intention of a collective virtual observer
is measured by the parameter $q_R$.
It should be marked here that $\delta A$ and $A^{\conf}$ are not mutually
local.

This example clarifies the notion of a collective virtual observer; it shows
that (1) only a part of recipted information is interpreted as an information
sent by a collective virtual observer (i.e. its presense does not demolish in
any way the presense of real ones), (2) a process of intention of a collective
virtual observer is completely determined by real ones in interaction (i.e. a
collective virtual observer is a specific united state of real ones entering
an interactive computerographical psychoinformation system in the multi--user
mode), (3) a collective virtual observer communicates with real observers
being interpreted (at least, formally) as an independent one.

\Refs
\roster
\item" [1]" Juriev D., Octonions and binucular Mobilevision; hep-th/9401047.
\item" [2]" Rheingold H., Virtual reality. Summit, New York, Tokyo, 1991.
\item" [3]" Virtual realities. In "Visual computing". Ed. T.L.Kunii, Springer,
1992.
\item" [4]" Visualized flow. Ed. Y.Nakayama. Pergamon, Oxford, 1988.
\item" [5]" Scientific visualization of physical phenomena. Ed.
N.M.Patrikalakis, Springer, 1991.
\item" [6]" Scientific visualization: techniques and applications. Ed.
K.W.Brodlie, Springer, 1992.
\item" [7]" Juriev D.V., Quantum projective field theory: quantum--field
analogs of Euler formulas. Teor. Matem. Fiz. 92:1 (1992), 172-176 [in
Russian].
\item" [8]" Juriev D.V., Quantum projective field theory: quantum--field
analogs of Euler--Arnold equations in projective $G$-hypermultiplets. Teor.
Matem. Fiz. 98:2 (1994), 220-240 [in Russian].
\item" [9]" Gervais J.-L., Conformal invariant field theories in two
dimensions, critical systems and strings. In "Workshop on Unified String
Theories". Eds. M.Green, D.Gross, World Scientific, Singapore, 1986, 459-468.
\item"[10]" Mack G., Introduction to conformal invariant quantum field theory
in two and more dimensions. In "Nonperturbative quantum field theory" Eds.
G.t'Hooft et al., Plenum, New York, 1988.
\item"[11]" Nahm W., Conformal quantum field theory in two dimensions. World
Scientific, Singapore, 1992.
\item"[12]" Ro\v cek M., Representation theory of the nonlinear $\SU(2)$
algebra. Phys. Lett. B255 (1991), 554-557.
\item"[13]" Bychkov S.A., Juriev D.V., Fubini--Veneziano fields in quantum
projective field theory. Uspekhi Matem. Nauk 46:5 (1991), 167-168 [in
Russian].
\item"[14]" Bychkov S.A., Fubini--Veneziano fields in projective Verma
quasibundles. Uspekhi Matem. Nauk 47:4 (1992), 187-188 [in Russian].
\item"[15]" Juriev D., An explicit form of the vertex operators in
two--dimensional $\slt$--invariant field theory. Lett. Math. Phys. 21 (1991),
113-115.
\item"[16]" Juriev D., The explicit form of the vertex operator fields in
two--dimensional quantum $\slt$--invariant field theory. Lett. Math. Phys. 22
(1991), 141-144.
\item"[17]" Juriev D.V., Classification of vertex operators in
two--dimensional quantum $\slt$--invariant field theory. Teor. Matem. Fiz.
86:3 (1991), 338-343 [in Russian].
\item"[18]" Bychkov S.A., Juriev D.V., Three algebraic structures of quantum
projective field theory. Teor. Matem. Fiz. 97:3 (1993), 336-347 [in Russian].
\item"[19]" Mi\v s\v cenko A.S., Fomenko A.T., Euler equations on finite
dimensional Lie groups. Isvestiya AN SSSR, Ser. Math. 12 (1978), 371-390 [in
Russian].
\item"[20]" Zakharov V.E., Manakov S.V., Novikov S.P., Pitaevskii L.P.,
Soliton theory: inverse scattering problem method. Moscow, Nauka, 1980 [in
Russian].
\item"[21]" Solitons. Eds. Bullough R.K., Caudrey P.L., Springer, 1980.
\item"[22]" Faddeev L.D., Takhtajan L.A., Hamiltonian approach to soliton
theory. Moscow, Nauka, 1988 [in Russian].
\item"[23]" Saati T.L., Speculating on the future of Mathematics. Appl. Math.
Lett. 1 (1988), 79-82.
\item"[24]" Beltrametti E.G., Cassinelli G., The logic of quantum mechanics.
Encycl. Math. Appl. 15, Addison--Wesley Publ., London, 1981.
\item"[25]" Juriev D.V., Complex projective geometry and quantum projective
field theory. Teor. Matem. Fiz. (1994) [in Russian].
\item"[26]" Juriev D., Algebraic structures of the quantum projective
($\slt$--invariant field theory): local projective field algebra $\Dscr(\slt)$.
J. Math. Phys. 33 (1991), 1153-1157.
\item"[27]" Juriev D., Alegbraic structures of quantum projective field theory
related to fusion and braiding. Hidden additive weight. J. Math. Phys. 35
(1994), 3368-3379.
\item"[28]" Bychkov S.A., Juriev D.V., Plotnikov S.V., Skladen' of Verma
modules over Lie algebra $\slt$ and hidden $\slth$--symmetries in quantum
projective field theory. Uspekhi Matem. Nauk 47:3 (1992), 155 [in Russian].
\item"[29]" Juriev D., Noncommutative geometry, chiral anomaly in quantum
projective ($\slt$--invariant) field theory and $\jlt$--invariance. J. Math.
Phys. 33 (1992), 2819-2822; 34 (1993), 1615.
\item"[30]" Juriev D., Remarks on nonassociative structures in quantum
projective field theory: the central extension $\jlt$ of the double
$\slt+\slt$ of the simple Lie algebra $\slt$ and related topics. Acta Appl.
Math. (1994).
\item"[31]" Juriev D., Algebraic structures of the quantum projective
($\slt$--invariant) field theory: the commutative version of E.Cartan's
exterior differential calculus. J. Math. Phys. 33 (1992), 3112-3116.
\item"[32]" Juriev D.V., QPFT--operator algebras and commutative exterior
differential calculus. Teor. Matem. Fiz. 93:1 (1992), 32-38 [in Russian].
\endroster
\endRefs
\enddocument